# The importance of overcoming MOVPE surface evolution instabilities for >1.3 μm metamorphic lasers on GaAs

*Enrica E. Mura,\*,† Agnieszka M. Gocalinska,† Megan O'Brien, † Ruggero Loi,†, § Gediminas Juska, † Stefano T. Moroni, † James O'Callaghan, † Miryam Arredondo,¥ Brian Corbett, † and Emanuele Pelucchi†*

† Tyndall National Institute, "Lee Maltings", University College Cork, Cork, Ireland

§ X-Celeprint Limited, Lee Maltings, Dyke Parade, Cork, Ireland

¥ School of Mathematics and Physics, Queen's University Belfast, Belfast, BT7 1NN, United Kingdom

**ABSTRACT**

We investigated and demonstrated a 1.3 µm-band laser grown by metalorganic vapour phase epitaxy (MOVPE) on a specially engineered metamorphic parabolic graded $In_xGa_{1-x}As$ buffer and epitaxial structure on a GaAs substrate. Bottom and upper cladding layers were built as a combination of AlInGaAs and InGaP alloys in a superlattice sequence. This was implemented to overcome (previously unreported) detrimental surface epitaxial dynamics and instabilities: when single alloys are utilised to achieve thick layers on metamorphic structures, surface instabilities induce defect generation. This has represented a historically limiting factor for metamorphic lasers by MOVPE. We describe a number of alternative strategies to achieve smooth surface morphology to obtain efficient compressively strained $In_{0.4}Ga_{0.6}As$ quantum wells in the active layer. The resulting lasers exhibited low lasing threshold with total slope efficiency of 0.34 W/A for a 500 µm long ridge waveguide device. The emission wavelength is extended as far as 1360 nm.

**INTRODUCTION**

GaAs-based lasers represent (together with "on silicon" approaches[1, 2]) an affordable and versatile alternative to conventional structures based on InP substrates for data communications over optical fibre in the 1.3–1.55 µm wavelength range. InP based devices have disadvantages not only due to substrate costs and restricted wafer size availability, but also on several other fronts. For example, the small material bandgap offsets and consequent electron overflow lead to degradation of several device characteristics even at moderate temperature operation. All this translates to poor characteristic temperature ($T_0$) in traditional 1.3 µm InGaAsP lasers, with $T_0$ of about 60 K,[3] while AlInGaAs based approaches only moderately improve performances.[4, 5] Even with the improved $T_0$, InP based diode lasers mostly require thermoelectric cooling in practical applications.[6]

The advantage of GaAs based lasers derives largely from the exploitation of lattice-matched Al containing layers for better carrier (higher bandgap offset) and optical (higher refractive index mismatch) confinements, and from compressively strained InGaAs quantum wells (QW) for high gain, all leading to better performing devices at high temperatures.[7, 8, 9] InGaAs QW lasers in the range of 900–1200 nm have been industrial products for a long time, showing excellent laser performance and stability. However, the longest lasing wavelength available to date is about ~1.26 µm,[10] due to limitations associated with the highly strained layers, this excluding InP metamorphic approaches.[11] Historically, several alternative routes have been investigated to achieve longer wavelength operation on GaAs: a) novel pseudomorphic strained QWs, b) quantum dot based active layers and c) metamorphic designs. Considering the first two approaches, for the 1.3 µm range significant research efforts, especially in terms of ultralow threshold current densities, have been made with InAs quantum dots (QDs)[12] and InGaNAs QW,[13] while excellent results have also been achieved with InGaAsN(Sb) dilute nitride alloy QWs[14] and In(Ga)As QDs.[15] Viable commercial operation, on the other hand, is limited to just below 1.3 µm,[16] and extension of the QD design on GaAs to longer wavelengths is considered a noteworthy challenge.

Metamorphic growth techniques to develop GaAs-based semiconductor lasers operating at 1.3 and 1.55 µm wavelengths have been broadly previously investigated.[17] Metamorphic growth generates a virtual wafer by controlled strain relaxation. For example, by depositing an $In_xGa_{1-x}As$ metamorphic buffer layer (MBL) on a GaAs substrate, heterostructures can be subsequently grown with a lattice constant matching to the in-plane lattice parameter of the grading, which is intermediate between that of GaAs and InP, maximising the advantage of GaAs substrates.[18] The challenge is to design a MBL capable of confining misfit dislocations and suppressing threading dislocations. Historically, the struggle to manage strain and defects dynamics translated into difficulties in achieving reliable working devices. Indeed, since 1994 when the first metamorphic InGaAs/GaAs telecom laser was reported by Uchida et al.,[19] only a few working lasers were reported reaching beyond 1.3 µm telecom wavelength, all fabricated using molecular beam epitaxy (MBE),[20, 21] while metalorganic vapour phase epitaxy (MOVPE) surprisingly failed in obtaining meaningful results. However, using MOVPE Nakao et al.[22] recently achieved 1.3 µm operation with native QWs photoluminescence at 1.27 µm. It is worth stressing that no explanation nor discussion is present in the literature on the limiting factors associated with MOVPE growth of such structures, nor how to overcome them.

We note briefly here that "on silicon" metamorphic approaches in combination of QD structures, in the double role of defect filters and active materials insensitive to dislocations, have shown recently a broad range of impressive results.[23] While acknowledging the potential of such structures, we will not further discuss the topic here as we concentrate on GaAs based structures and their challenges by MOVPE. Obviously our findings will indeed be useful to the broader community, as well as for "on silicon" approaches.

This paper discusses the growth and fabrication of metamorphic InGaAs multi quantum well (MQW) lasers emitting beyond 1.3 µm (with the potential for extension to >1.5µm) exploiting doped graded InGaAs metamorphic buffer layers on GaAs substrates. Our results crucially rely on understanding the specific MOVPE surface dynamics for the formation of the cladding and compressively strained MQW layers, which enforces the material choices and combinations. We demonstrate an adjustable and reproducible approach for a working laser device emitting in the datacom range. Electro-optical measurements of the metamorphic laser device are presented to demonstrate successful operation and potential for further improvements.

**METHODOLOGY**

All the epitaxial samples discussed here were grown in a MOVPE commercial horizontal reactor at low pressure (80 mbar) with purified $N_2$ as carrier gas.[24] Trimethyl-III organics (TMIn, TMGa, TMAl), arsine ($AsH_3$) and phosphine ($PH_3$) were used as group III and V precursors, (TMSb) was used as a surfactant during cladding layer growth; diethylzinc (DEZn) and disilane ($Si_2H_6$) were used as p- and n-dopant sources, respectively. The epitaxial growths were performed mainly on (100) GaAs misoriented substrates 0.2° and 6° towards [111]A. The surface morphology and structural defect formation were analysed by atomic force microscopy (AFM) in tapping non-contact mode at room temperature and in air. The assessment of composition and the strain in the layers was made according to measurements of reciprocal space maps (RSM) obtained by high resolution X-ray diffraction measurements (HRXRD). Measurements were done in a symmetric (004) and in two asymmetric (224 and -2-24) reflections with sample positioned at 0°, 90°, 180° and 270° with respect to its main crystallographic axes. To acquire the composition of quaternary alloys, being the generic quaternary alloy $A_xB_{1-x}C_yD_{1-y}$ specified by two variables, Vegard's law was combined with the energy gap obtained by photoluminescence measurements.

Transmission electron microscope (TEM) characterisation was performed using a Thermo Fisher Talos F200X G2 fitted with the Super-X energy dispersive X-ray spectrometer (EDS) operated at 200 kV using high angle annular dark field (HAADF) scanning TEM (STEM) and EDS.

The structure of the target laser device developed in this paper starts with a homoepitaxial GaAs 100 nm thick buffer followed by a metamorphic graded n+ $In_xGa_{1-x}As$ layer (~0<x<18%), 1 μm thick. Then, a laser waveguide structure containing n-doped lower cladding, MQW active, p-doped upper cladding and contact layers was grown, all pseudomorphic to the in-plane lattice parameter at the end of the grading. The cladding layers were grown with signature superlattice structures built as a combination of AlInGaAs and InGaP alloys. The lower and upper cladding layers were Si-doped and Zn-doped, with concentration of ~$1\times10^{18}$ cm$^{-3}$ and ~$8\times10^{17}$ cm$^{-3}$ respectively. The active region of a separate confinement heterostructure (SCH) consisted of three compressively strained nominal $In_{0.40}Ga_{0.60}As$ quantum wells 7 nm thick, sandwiched between nominal $In_{0.13}Ga_{0.87}As$ and $Al_{0.12}In_{0.14}Ga_{0.74}As$ barrier layers, and preceded by 5 nm of GaAs used as control interface layer (CIL). Finally, a 100 nm thick InGaAs top p-contact layer with Zn doping of $1\times10^{19}$ cm$^{-3}$ was grown. The growth temperature was kept at 740 °C (thermocouple reading, which corresponds to an estimated real surface temperature of ~650 °C) throughout the laser structure except in the active region, where it was decreased down to 580 °C (or estimated real ~ 540 °C) to avoid 3D nanostructure formation, as discussed later.

The complete epitaxial structure of the full laser is sketched in **Figure 1a.** The $In_xGa_{1-x}As$ MBL was deposited on GaAs buffer following a single parabolic grading profile. The design, based on a model discussed in Ref.,[25] was optimized as reported in Ref.[26]. The STEM HAADF image of the full laser structure (**Figure 1b**) confirmed that most of the threading dislocation network was buried down close to the GaAs substrate. The in-plane lattice parameter in representative calibration samples was equivalent to a fully relaxed $In_{0.14}Ga_{0.76}As$, (by HRXRD); the residual parallel strain was -0.0030%. At the top of the MBL the surface showed a cross-hatch pattern, a common characteristic in metamorphic structures. The ridges aligned along [110] and [1-10] directions lead to an asymmetry of the roughness in the two <110> directions, with a higher root mean square (RMS) value in [110] direction than the [1-10] direction. The RMS values (evaluated from an AFM scan size area of 50 ×50 μm$^2$, not shown here) were typically of ~5 nm along the [110] direction, while were ~3 nm along the [1-10] direction, for both 0.2°A and 6°A substrate misorientations studied.

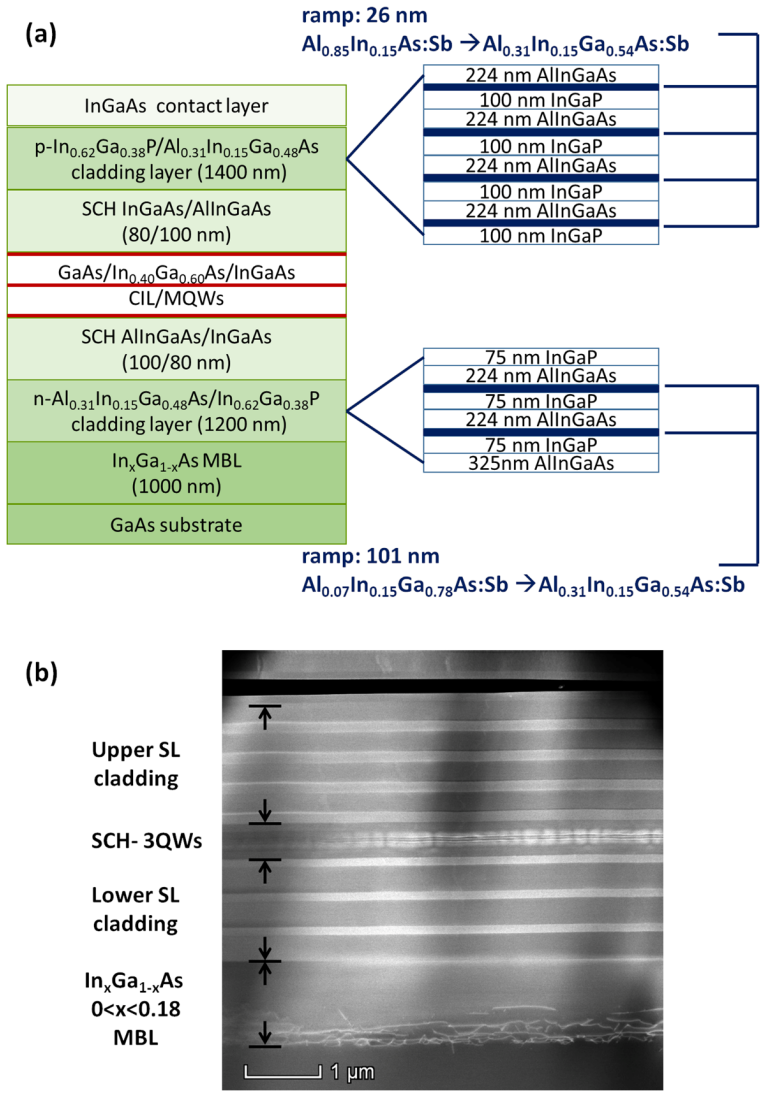

**Figure 1.** (a) Schematic of the epitaxial laser structure with a zoom in the cladding layers. (b) Cross-sectional STEM HAADF image of the full laser grown on GaAs 0.2°A misoriented substrate. The apparent waviness is induced by the mechanical sample preparation.

**RESULTS AND DISCUSSION**

The surface organization and roughness behaviour is preserved in the subsequent cladding layers deposited onto the MBL by using an AlInGaAs-InGaP superlattice (SL). AlInGaAs and InGaP represent two suitable alloys for the cladding selection in order to maintain efficient optical field confinement and waveguiding, considering a 1.3 μm emission in a SCH multi-QW configuration for lasing. Indeed, for the quaternary AlInGaAs layer, the Al composition can be significantly varied while keeping lattice matching to the $In_xGa_{1-x}As$ MBL. Moreover, the AlInGaAs material system provides simplicity to a metamorphic laser design, as varying the aluminium content allows both strong QW confinement when used for barriers, and engineering of the SCH to enhance the optical confinement, hence reducing the material gain at threshold, as recently discussed[18]. On the other hand, InGaP instead of AlInGaAs, is often used as upper cladding in asymmetric laser structure design because, when wet etched for example, there are no complications induced by oxidation.[27] We observed that bulk AlInGaAs or InGaP when grown pseudomorphically on top of an MBL, shows a strong correlation between the epilayer thickness, surface roughness and subsequent defect generation. When layers were grown with thicknesses typical for the cladding and overall structure (i.e. >>1 micron), inevitably a strong enhancement of RMS values emerged and v-shaped surface defects appeared, with an overall unacceptable degradation of layer quality. It was also observed that for thicknesses less than 300 nm, both AlInGaAs and the InGaP alloys had RMS values of the same order of magnitude or close enough to those of the MBL. Hence, to avoid the surface instabilities which appeared with increased layer thicknesses when a single alloy was utilised, the strategy was to combine the two alloys in the lower cladding barrier, alternating them while keeping the thickness for each layer around 300 nm for the AlInAs layer and beyond for the InGaP.

**Figure 2** (left side) highlights how the SL approach out-performs a single alloy in terms of RMS value and reduced peak to valley distance. Increasing the thickness of the AlInGaAs layer led to a higher RMS value, from ~11 nm for the 300 nm thick layer to ~23 nm for the AlInGaAs layer 1400 nm thick preceded by an InGaAs strain balancing layer. Indeed, the attempted addition of a strain balancing layer before AlInGaAs layer deposition did not improve the surface roughness, despite showing promising results in previous works (Ref.26). Furthermore, the AlInGaAs compound showed a deteriorated surface with the subsequent deposition of other similar or thicker AlInGaAs layers. For example, we observed an RMS value of ~16 nm (evaluated by an AFM scan size 10×10 μm$^2$, image not shown here) for a sample grown with $Al_{0.13}In_{0.17}Ga_{0.7}As$ 1000 nm followed 50 nm of $Al_{0.40}In_{0.17}Ga_{0.43}As$ pseudomorphic to the MBL. Similarly, the InGaP analysis revealed the same trend but amplified: i.e. the degradation of the surface was strictly proportional to the layer thickness and was accompanied by deep trenches on the surface. The alternation of the peaks and valley on the surface, typical of the cross-hatch pattern of the MBL, increased several orders of magnitude with the thickness. The maximum "trench-like" depths were around 15 nm for the MBL, while, when 1400 nm of InGaP was deposited directly on the metamorphic substrate, grooves appeared with depth reaching, e.g., more than 150 nm (a sample grown on 0.2° towards [111]A substrate) and the RMS value exceeded 30 nm. V-shaped defects started to show with increased layer thickness, and seemed to be uncorrelated to the original MBL relaxation defects close to the interface with GaAs, but associated with the building up of surface roughness and instabilities; they certainly originated close enough to the sample top surface and away from the defected region (not shown).

In **Figure 2** (right side) we show the AFM signal amplitude of a lower cladding SL consisting of 250 nm of $Al_{0.31}In_{0.15}Ga_{0.54}As(Sb)$ pseudomorphic (PM) followed by 50 nm $In_{0.62}Ga_{0.38}P(Sb)$ tensile strained, repeated 5 times for a total thickness of 1500 nm. The RMS value remained around 6 nm, and no new defects appeared. In the SL design, the resulting InGaP surface improved by adding some tensile strain directly into the InGaP layer itself. In the final structures, both AlInGaAs and InGaP layers were grown with TMSb as a surfactant,[28, 29] which proved effective in helping to maintain a small RMS.[30] In the Supporting Information (**Figure S1**) we report AFM images (height, signal amplitude and cross-sectional profile) of all the sample mentioned in **Figure 2**.

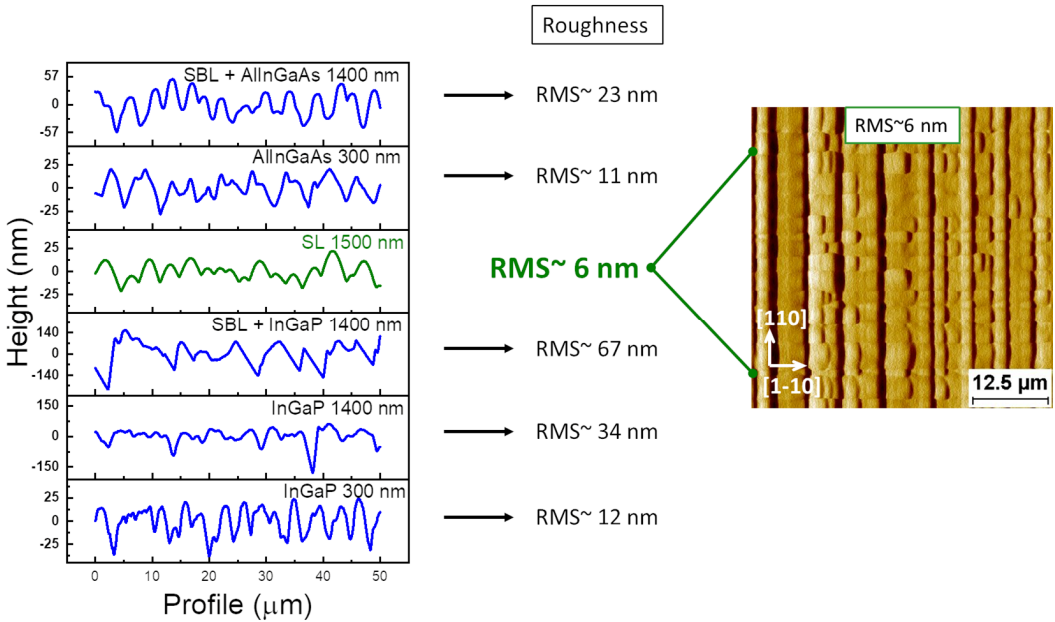

**Figure 2**. Representative surface morphology of samples with various cladding designs. Left side: comparison between different cladding and the combined superlattice structures in terms of AFM cross-sectional profiles, thickness and RMS value. Right side: AFM signal amplitude of lower cladding SL with layer sequence: 250 nm of $Al_{0.31}In_{0.15}Ga_{0.54}As(Sb)$ followed by 50 nm $In_{0.62}Ga_{0.38}P(Sb)$ tensile strained, repeated 5 times for a total thickness of 1500 nm. All samples presented here were grown on perfectly oriented GaAs substrates or misoriented of 0.2° towards [111]A.

When the upper cladding had a SL design, we observed that not only did the increase of the thickness contribute to the degradation of the surface roughness, but also that the order of the layer deposition needed to be controlled (surprisingly). As noted in case of the n-cladding, the deterioration of the surface is linked to the increase of layer thickness; the same happened with p-cladding. We also observed that increasing the thickness of a single alloy leads to an increase in defects. According to the proposed SCH scheme, the active part ends with a barrier of $Al_{0.12}In_{0.15}Ga_{0.73}As$. The subsequent deposition of $Al_{0.30}In_{0.15}Ga_{0.55}As$, 250 nm thick, as the first layer of the p-cladding, resulted in visible defect lines on the surface. In Figure 3 the phenomenon can be observed for a SL upper cladding structure grown at the end of the active part, with the following layer sequence: 250 nm of $Al_{0.31}In_{0.15}Ga_{0.54}As(Sb)$ followed by 50 nm $In_{0.62}Ga_{0.38}P$ tensile strained, repeated 2 times for a total thickness of 600 nm. The sample shown in Figure 3 was grown on 6°-off GaAs substrate. The degradation effect was more intense in the sample grown using a 0.2° -off GaAs substrate (please refer to Supporting Information Figure S2), where the RMS exceeded 35 nm. Reversing the deposition sequence of the upper p-cladding SL sequence, i.e. InGaP first and AlInGaAs after, restored a surface free from visible threading dislocations.

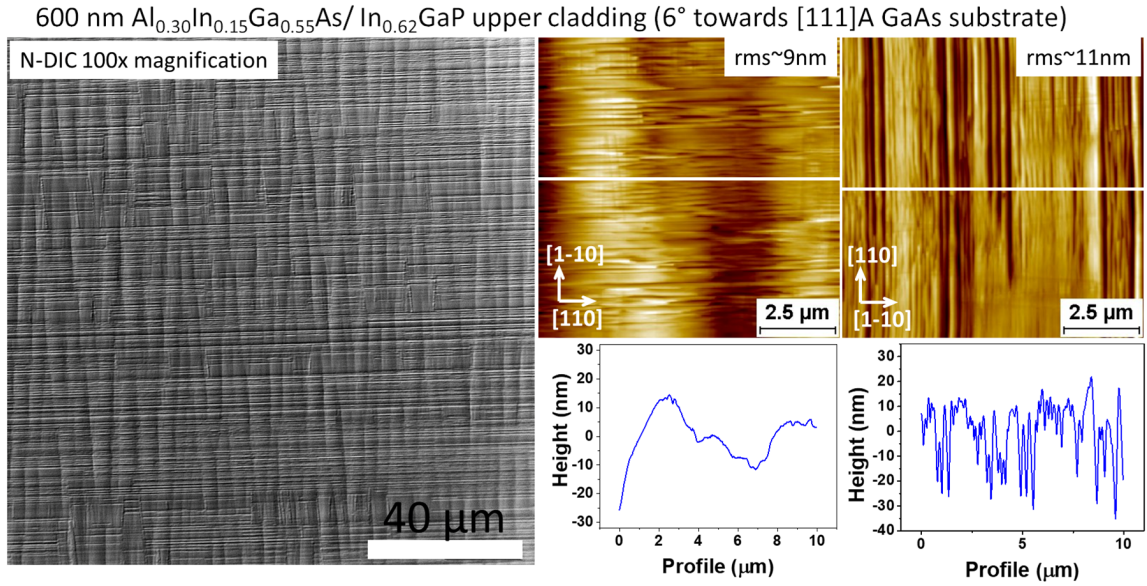

**Figure 3**. (Nomarski) Differential Interference Contrast and AFM (signal height and cross-sectional profile along the two [110] and [1-10] directions) images of SL upper cladding grown at the end of active part, with following layer sequence: 250 nm of $Al_{0.31}In_{0.15}Ga_{0.54}As(Sb)$ followed by 50 nm $In_{0.62}Ga_{0.38}P$ tensile strained, repeated 2 times for a total thickness of 600 nm. The defect line structures are linked to the growth sequence (see text).

The cladding layer sequence was also optimized (returning to the structure sketched in **Figure 1a**) by linearly ramping the composition of the AlInGaAs layer from $Al_{0.07}In_{0.15}Ga_{0.78}As$ to $Al_{0.31}In_{0.15}Ga_{0.54}As$ over 101 nm at the InGaP/AlInGaAs interface in the lower (n-doped) part. The p-cladding was modified at the InGaP/AlInGaAs interface where the percentage of gallium was linearly graded from ~0% to 54% and the aluminium from 85% to 31%, keeping the indium content constant, over a thickness of 26 nm. The intent was to facilitate a better carrier transport toward the active part by improving the band alignment between AlInGaAs and InGaP in the conduction and in the valence bands, as conventionally done in VCSELs designs.[31] Also, the number of interfaces was reduced to 4, increasing both AlInGaAs and InGaP thicknesses.

However, we did this keeping the thickness for each layer below or around 300 nm, as this value was recognized as the limit for a "safe" defects free surface.

The growth temperature was kept constant at 740°C (thermocouple) during the full lower cladding deposition and then was gradually decreased at the end of the $Al_{0.12}In_{0.14}Ga_{0.74}As$ layer in the SCH part, for QWs growth at 580°C. The heavy compressive strain in QWs and the residual strain in the metamorphic buffer layer (in combination with the surface step bunched ordering) promoted 3D feature formation under certain growth temperatures and for certain percentage of indium in the QWs. To avoid and control the 3D nanostructuring, we inserted a thin GaAs layer before the QW region as a controlled interface layer. Five nm of GaAs were sufficient to delay the nanostructure formation process, guaranteeing 2D growth without affecting the active layer optical emission. We systematically investigated the range of growth temperatures and indium content in the active layers which were both 3D-nanostructured and defect-free, while maintaining the emission wavelengths of interest. A range between 540 °C and 650 °C (thermocouple) represents a solid and useful interval where we could grow 2 InGaAs QWs with up to 40% of indium, whereas for three InGaAs QWs with the same 40% of indium the temperature was decreased to 580 °C. HAADF images confirmed the absence of defects and 3D features in the active part (**Figure 1, Figure 4a, 4b**). Photoluminescence (PL) at room temperature on calibration samples showed active layer emission at ~1360 nm (**Figure 4c**).

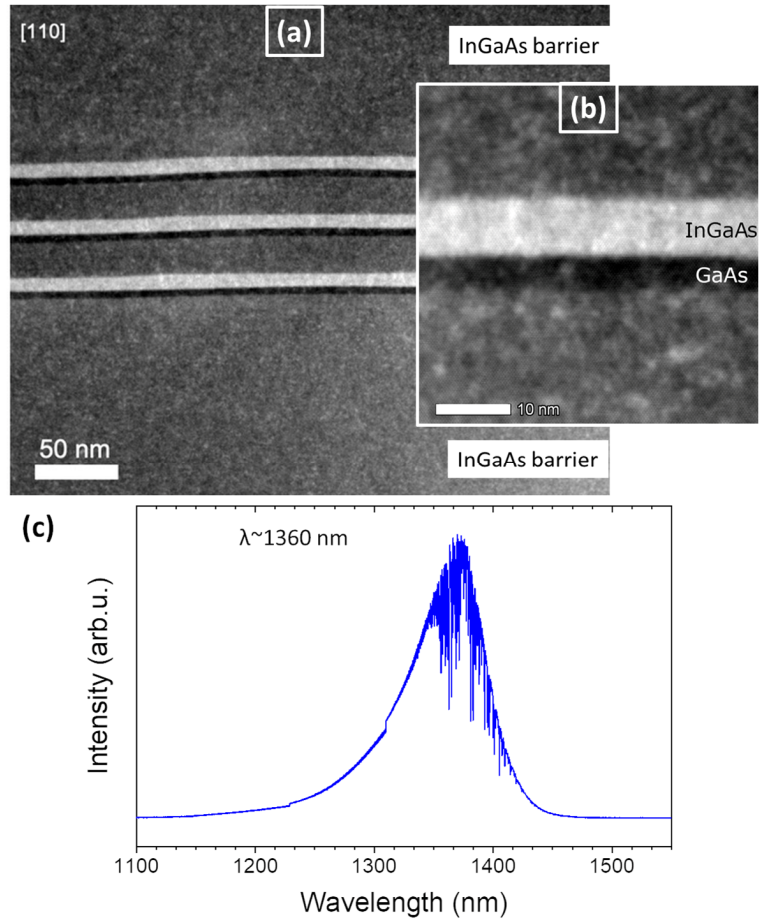

**Figure 4**. (a) and (b) HAADF STEM images zoomed-in the laser active part, composed by three $In_{0.40}Ga_{0.60}As$ QWs, preceded by GaAs CIL layer and separated by $In_{0.13}Ga_{0.83}As$ barrier. (c) The room temperature PL spectrum showing emission at ~1360 nm from the $In_{0.40}Ga_{0.60}As$ MQWs structure grown at 580 °C (the spectral modulations are due to the absorption of atmospheric gases present in the optical path of the measurement set-up). Samples grown on GaAs misoriented substrate 0.2° towards [111]A.

Based on these epitaxial structures a number of ridge injected laser configurations were fabricated with widths ranging from 4 µm to 20 µm using Ti/Au and AuGeNi as the metallisations for the p- and n-type contacts respectively. There was no significant difference in the performance for the lasers realised on the 0.2° and 6°-off substrates. A mode confinement factor of 3% was calculated for the 3QW structure. The light-current (LI) characteristics of devices from the 3-QW structure for a range of cavity lengths with uncoated facets were measured in pulsed and continuous-wave (CW) mode as a function of temperature. **Figure 5a** shows the CW single facet output power characteristics from a 500 µm long 4.5 µm wide ridge laser. The device reaches a total (both facets) slope efficiency of 0.35 W/A. The device operates in CW up to 65 °C limited by self-heating (see below). The resistance of the device is 5.5 Ω confirming the excellent carrier transport across the interfaces at the superlattices due to the composition grading at the interfaces. The temperature dependence of the threshold current, $I_{th}$, is characterised through the $T_0$ value given by $T_0 = dT/d\ln I_{th}$. In CW operation, for the 500 µm long device, a value of 60 K is obtained (inset **Figure 5a**). **Figure 5b** shows the pulsed (pulse length 200 ns, duty cycle 0.1 %) LI characteristic from a 900 µm long and 20 µm wide device. The threshold current density for the laser is 1.7 kA/cm$^2$ with a $T_0$ value of 190 K measured. This shows that the QW region allows high temperature operation. The threshold current does not scale with ridge width (in this case) due to the quality of the cleaved facet affected by the strain associated with the metamorphic buffer.

The CW spectrum at 20 °C of the 300 µm x 4.5 µm laser is shown in **Figure 6**. The emission wavelength reached was as long as 1360 nm. There was a 30 nm shift in the peak wavelength for a 70 mA change in the current (80–150 mA). Since the temperature dependence of the bandgap was estimated to be 0.6 nm/K, the junction temperature increase was estimated to be 50 K. This indicates a very high thermal resistance of >300 K/W which is attributed to high thermal resistance of and thus poor heat dissipation through the SL and MBL. This current-induced heating is limiting the CW thermal performance of these lasers. A final issue to be noted is the measured transverse far-field of the lasers with wider ridge width as shown in **Figure 7**. There, the intensity as a function of pulsed injected current is shown with a full width at half maximum of 60°. A discontinuity in the far-field profile is observed at an angle -40° and this is associated with the coupling of waveguide energy to the high index MBL and GaAs substrate. The far-field and mode coupling is confirmed by simulations of the mode structure where the substrate coupling reduces with reducing ridge width. Despite the thermal and optical issues, the laser performance is very promising, demonstrating high gain at high temperature, low electrical resistance and extended wavelength operation. Many options exist for improving the performance through mode engineering and structures for thermal dissipation.

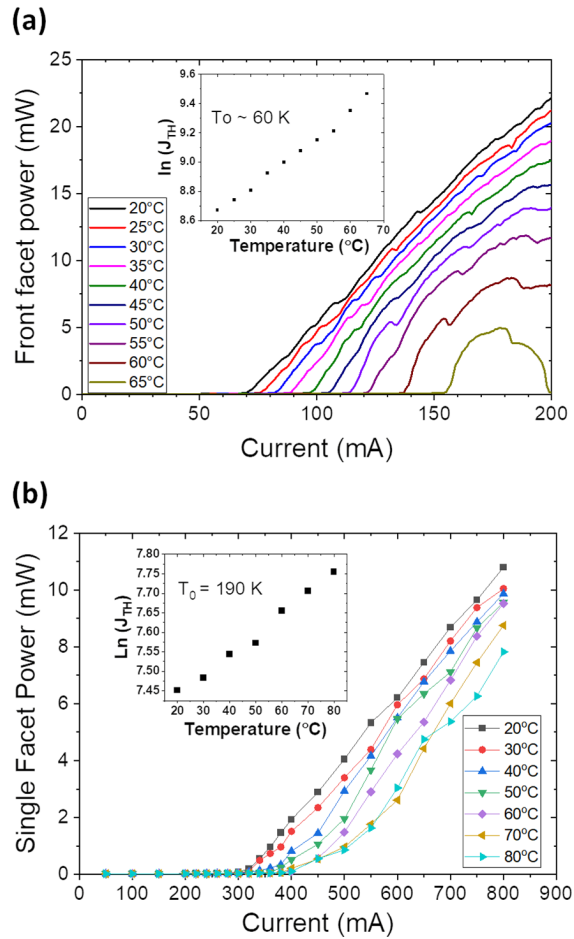

**Figure 5**. Temperature dependent L-I characteristics under (a) CW operation for 500 μm long x 4.5 μm wide ridge laser, (b) under 200 ns pulse excitation for 900 μm long x 20 μm wide ridge laser.

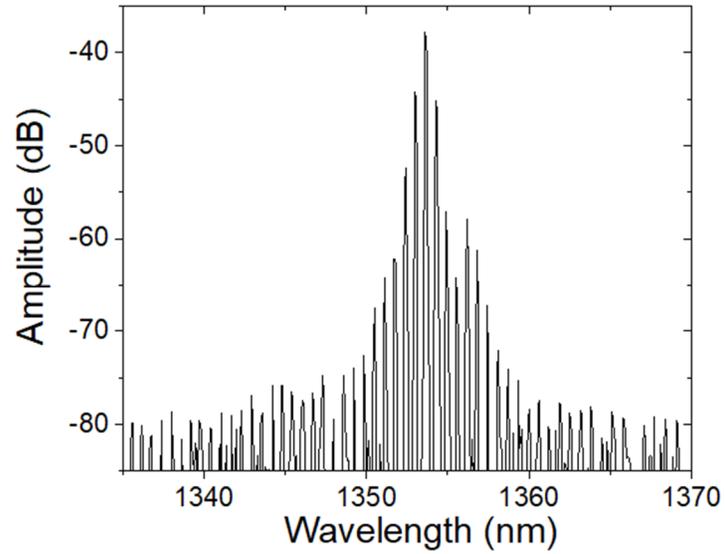

**Figure 6**. CW lasing spectrum from a 300 μm long x 4.5 μm wide ridge laser driven at 120 mA.

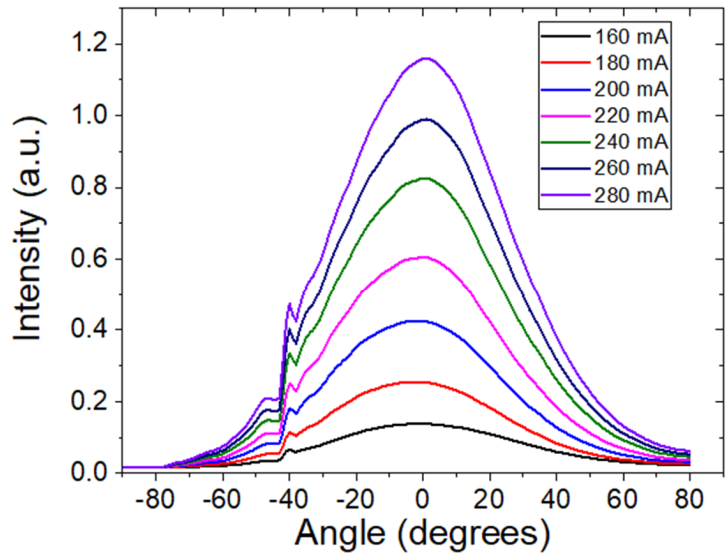

**Figure 7**. Transverse far-field and evolution with pulsed current from a 350 μm long x 10 μm wide ridge laser.

**CONCLUSIONS**

We have demonstrated an O-band laser structure on an engineered metamorphic graded InGaAs buffer grown by MOVPE on a GaAs substrate. We discussed that the culprit inhibiting MOVPE metamorphic lasers > 1.3 µm is linked to previously unreported surface instabilities. Strain relaxation defects at the substrate interface are typically blamed for metamorphic device failures. In this case the catastrophic surface roughening appears to be largely independent from the buffer morphology, but tight to further surface dynamics. We presented a two-alloy superlattice cladding layer structure as a novel solution to control the epitaxial surface instabilities. We determined a range of growth temperatures and indium content in the active layer, which provided flat and defects free epitaxy while resulting in the emission of interest. The method combined a relatively low growth temperature with the insertion of a thin GaAs layer deposited before the QWs. The 4.5 µm wide and 500 µm long ridge lasers showed a low electric resistance with a total slope efficiency of 0.35 W/A. Under pulsed operation a $T_0$ of 190 K was measured. Our results represent a solution to long standing problems in the material science community on the MOVPE growth of thick metamorphic structures, and highlights the potential of the GaAs platform as replacement for InP-based laser structures. The achieved control over the material issues opens routes to "elastic" designs targeting specific wavelengths and extending the spectral span of previously available structures, possibly deeper into the infrared region.

ASSOCIATED CONTENT

**Supporting Information**

Additional morphology surface details of cladding designs are shown in Supporting Information.

Figure S1. AFM signal amplitudes, height, and cross-sectional profile

Figure S2. Differential Interference Contrast and AFM

AUTHOR INFORMATION

**Corresponding Author**

**Enrica Mura**-Tyndall National Institute, University College Cork, Cork, Ireland

* e-mail enrica.mura@tyndall.ie, Phone: +353 21 420 6195

**Author Contributions**

   The manuscript was written through contributions of all authors. All authors have given approval to the final version of the manuscript.

**Notes**

The authors declare no competing financial interest.

**Funding Sources**

This research was enabled by the Irish Higher Education Authority Program for Research in Third Level Institutions (2007-2011) via the INSPIRE Programme and partly by Science Foundation

Ireland under the IPIC award 12/RC/2276, 12/RC/2276_P2, and also grant 15/IA/2864. Also by the Irish Research Council under grant EPSPG/2014/35.

ACKNOWLEDGMENT

The authors would like to thank Dr. Silviu Bogusevschi and Prof. E. O'Reilly for useful discussions and providing insight and in alloy band alignment.

surfactant paradigm for phosphides as partially incorrect (more comprehensive study will be the object of future reports).

[31] Zhou, P., Cheng, J., Schaus, C.F., Sun, S.Z., Zheng, K., Armour, E., Hains, C., Hsin, W., Myers, D.R., Vawter, G.A. Low series resistance high-efficiency GaAs/AlGaAs vertical-cavity surface-emitting lasers with continuously graded mirrors grown by MOCVD. *IEEE Photon. Technol. Lett.* 1991, *3*, 591-593.

# The importance of overcoming MOVPE surface evolution instabilities for >1.3 μm metamorphic lasers on GaAs

*Enrica E. Mura,\*,† Agnieszka M. Gocalinska,† Megan O'Brien, † Ruggero Loi,†, § Gediminas Juska, † Stefano T. Moroni, † James O'Callaghan, †  Miryam Arredondo,¥ Brian Corbett, † and Emanuele Pelucchi†*

† Tyndall National Institute, "Lee Maltings", University College Cork, Cork, Ireland

§ X-Celeprint Limited, Lee Maltings, Dyke Parade, Cork, Ireland

¥ School of Mathematics and Physics, Queen's University Belfast, Belfast, BT7 1NN, United Kingdom

# Supporting Information

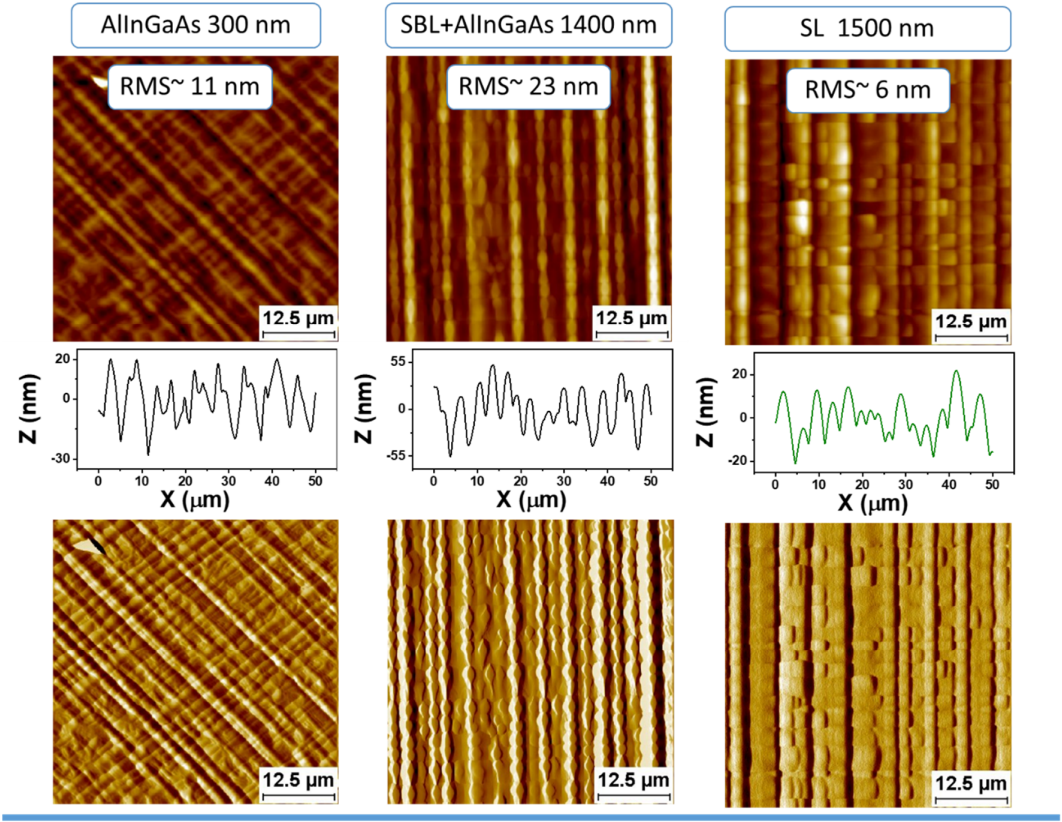
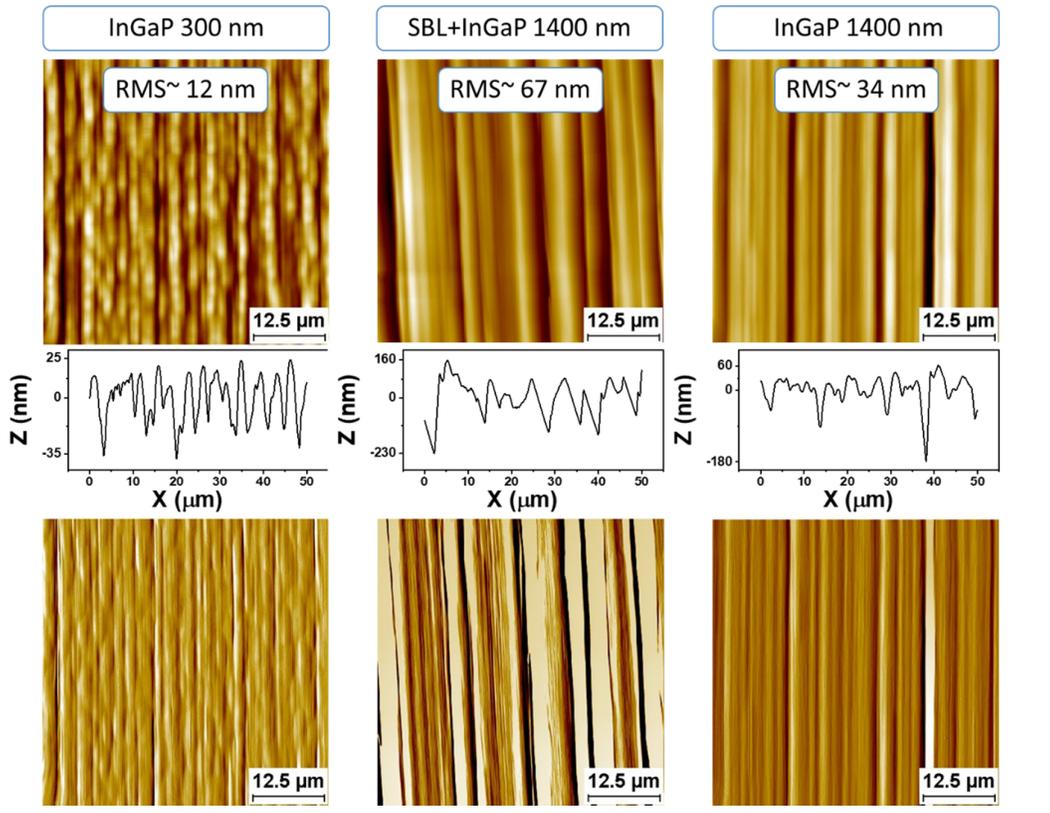

**Figure S1**. Surface morphologies (AFM signal amplitudes and height, Z in nm) of samples with various cladding designs. Comparison between different cladding and the combined superlattice SL structures in terms of AFM height and amplitude signal, cross-sectional profiles, thickness and RMS value. All samples presented here were grown on perfectly oriented GaAs substrates or misoriented of 0.2° towards [111]A. SBL indicates a strain balancing layer.

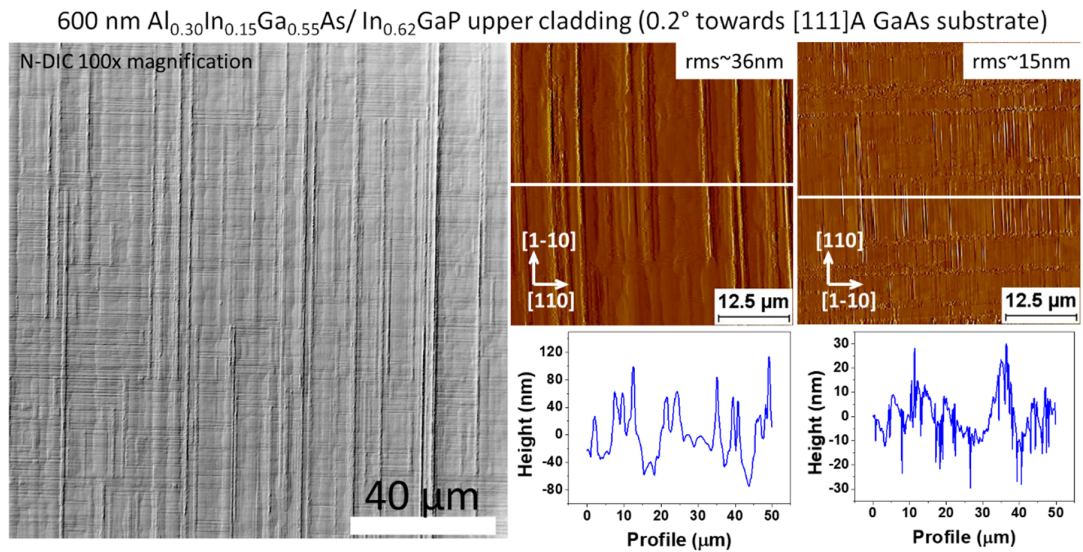

**Figure S2**. (Nomarski) Differential Interference Contrast and AFM (signal amplitude and cross-sectional profile) images of SL upper cladding grown at the end of active part, with following layer sequence: 250 nm of $Al_{0.31}In_{0.15}Ga_{0.54}As(Sb)$ followed by 50 nm $In_{0.62}Ga_{0.38}P$ tensile strained, repeated 2 times for a total thickness of 600 nm.